# Measurement bias: a structural perspective

Yijie Li [1,2,3], Wei Fan [1,2,3], Miao Zhang [1,2,3], Lili Liu [1,2,3], Jiangbo Bao [4,*], Yingjie Zheng [1,2,3,*]

[1] *Department of Epidemiology, Fudan University, Shanghai, China*

[2] *Key Laboratory of Health Technology Assessment, National Health Commission, Fudan University, Shanghai, China*

[3] *Key Laboratory of Public Health Safety, Ministry of Education, School of Public Health, Fudan University, Shanghai, China*

[4] *School of Nursing, Fudan University, Shanghai, China*

*Corresponding authors:

Yingjie Zheng, yjzheng@fudan.edu.cn, Department of Epidemiology, School of Public Health, Fudan University, 130th Dong'an Road, Shanghai, 200032, China; Jiangbo Bao, jbbao@fudan.edu.cn, School of Nursing, Fudan University, No.305 Fenglin Road, Shanghai 200032, China.

**Abstract**

The causal structure for measurement bias (MB) remains controversial. Aided by the Directed Acyclic Graph (DAG), this paper proposes a new structure for measuring one singleton variable whose MB arises in the selection of an imperfect I/O device-like measurement system. For effect estimation, however, an extra source of MB arises from any redundant association between a measured exposure and a measured outcome. The misclassification will be bidirectionally differential for a common outcome, unidirectionally differential for a causal relation, and non-differential for a common cause between the measured exposure and the measured outcome or a null effect. The measured exposure can actually affect the measured outcome, or vice versa. Reverse causality is a concept defined at the level of measurement. Our new DAGs have clarified the structures and mechanisms of MB.

Introduction

Scientists can only understand the world through measurement. With measurement, there would be measurement bias (MB) [1-5]; without measurement, we would know nothing about the world. This is a dilemma that we have to face, measuring something and producing MB.

In a source population, assuming there exists a property of an event (e.g. an exposure) we are interested in, we can apply a measurement system, such as observing, surveying, detecting, etc., to obtain its true value (denoted as a singleton variable, A). Unfortunately, we can only get its measured version (A*) instead. Generally, the measurement system ($M_{A*}$) is a combination of some critical elements, involving researchers, study subjects, instruments, methods, etc., throughout the whole measuring process in which it acts like a certain device with the true A input and the measured A* output. The difference between the values of A and A*, that is $U_{A*}=A*-A$, is known as the MB ($U_{A*}$), or misclassification bias if A is a categorical variable [6-9].

Measurement is anyway a causal issue [10-12]. Since the introduction of Directed Acyclic Graphs (DAGs) 30 years ago [13-16], it has been used widely in a range of areas [17-26], such as effect identification, selection of sufficient sets, mediation analysis, and collinearity, including MB [10-12,27-37]. Not surprisingly, these MB-DAGs were mainly proposed for dealing with the effect of an exposure over an outcome [30,31,37]; however, there have been insofar only few introductions of MB for one singleton variable. Among these introductions, a consensus has been reached that all factors other than A that affect the value of A* will result in MB. However, currently there are two major strategies differing in the treatment of MB. Hernán et al. used $U_{A*}$ to represent all the factors other than A that determined the value of A* [31], while Shahar et al. treated those factors separately [30,37]. The former approach could not provide us with the exact meaning of those factors, while the latter paralleling enumeration failed to capture the predominant features of the measurement system which had aimed at obtaining the A* value.

In addition, some other problems remain to be clarified. Firstly, the value of A can be assumed to be an unknown constant at the exact timing of measurement, thus if we obtain the value of A*, we will obtain the value of $U_{A*}$ accordingly. Furthermore, $U_{A*}$ is theoretically a mathematical difference, and there exists no place for it in causal reality. Thus, the arrows connecting $U_{A*}$ with other variables in the published MB-DAGs may be incorrect. Secondly, when studying the effect of an exposure (A) on an outcome (Y), we all know that one measured and known variable, A* (or Y*), will not affect the true value of another unmeasured and unknown variable, Y (or A); however, it is quite puzzling how this unknown variable, A (or Y), can directly affect another known variable, Y* (or A*), if no other variables intercept them. Thirdly, when the variables are discrete, their differential characteristics in the mechanism of misclassification is a critical concern; however, up to now, no causal structures have been presented to clarify these issues. Fourthly, since only one or several measurement systems will practically be selected, why this selection has never been depicted in any MB-DAGs?

## Methods

In this paper, we propose a new structure for measuring one singleton variable, and then extend it into clarifying the causal relationships with the illustration of several examples. Based on the measurement level, the DAGs can be helpful in clarifying the structures and mechanisms of MB.

## Results

### 1. A new MB-DAG for a singleton variable

Under specific spatiotemporal conditions, we can assume that the value of A is an unknown constant. Taking glucose tolerance test as an example, although the advance glucose intake for the individuals leads to an increase of this indicator, its true value at the time of measurement is fixed. Thus, we consider A as an exogenous variable. Unlike the A, the value of A* and $U_{A*}$ may vary with different measurement systems.

There are two factors affecting the value of A*. One should be the true value of A, and the other is the measurement system ($M_{A*}$), for the reason that there will be no MB without the implementation of this I/O device-like $M_{A*}$. Prior domain knowledge on the event or state of A* ($K_{A*}$) paves a way for our understanding and selection of the $M_{A*}$ [33]. The more we know about $K_{A*}$, the more likely we can propose many potential candidates $M_{A*}$ in view of their performance, cost, and feasibility. Among these candidates, only a relatively more scientific, rational and effective one will be generally selected and put into practice ($S_{A*}$). Furthermore, a more effective system will likely produce a more accurate value of A*. Based on these assumptions, we propose a new MB-DAG for measuring a singleton variable, A (Fig. 1).

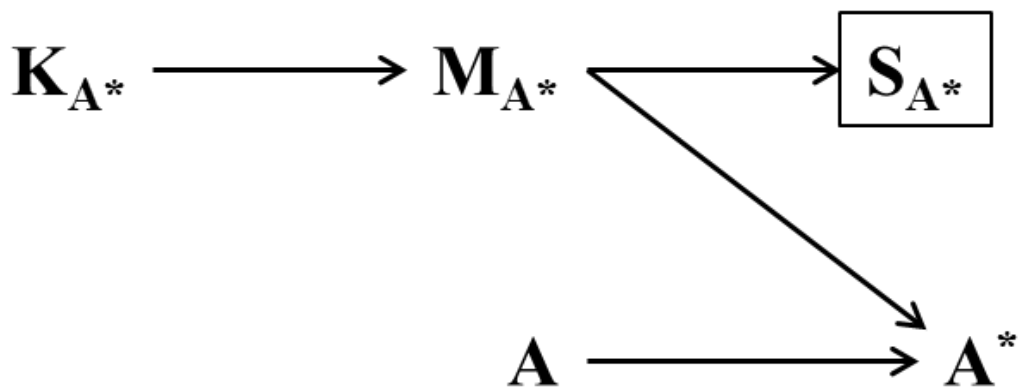

**Fig. 1 A directed acyclic graph for the measurement of a singleton variable (A)**

A: true value of a singleton variable; A*: measured version of A; $K_{A*}$: prior knowledge of A*; $M_{A*}$: candidate measurement systems for the value of A; $S_{A*}$: the measurement system selected into practice. The box around $S_{A*}$ means conditioning or limiting $M_{A*}$ to the actual measurement system.

We can assume that there exists one true-value measurement system (TVMS) for measuring the value of A. In Fig. 1, we can imagine two $M_{A*}$ are selected, one is the TVMS and the other is an actual-value measurement system (AVMS). Therefore, it seems that we were comparing the values obtained by AVMS with those by TVMS. If AVMS was the same as TVMS, we could expect to get a perfect copy of the value of A, i.e., A*=A. Unfortunately, even if there is one measurement system with the highest sensitivity and specificity, often referred as the "gold standard" in reality, it is unlikely to mirror the TVMS perfectly. Practically, we can only obtain the value of A* by AVMS through actual conditioning on $S_{A*}$, thus the path of $K_{A*} \rightarrow M_{A*} \rightarrow A*$ could not be completely blocked. As a result, MB will inevitably occur.

## 2. MB-DAGs for the effect of an exposure on an outcome

### 2.1. The substitution estimate for the effect: the independent and dependent mechanism of the measurement systems

When an exposure (A) and an outcome (Y) are measured respectively with their corresponding AVMS independently, where Y-related variables are named in the same way as those of A, we can only estimate the A*-Y* association, a surrogate estimate of the A-Y effect, for the reason that both A and Y are unknown in practice. Since the MB for a singleton variable often inevitably arises during the measurements on both A and Y which generally apply only one AVMS respectively, it could be expected that the A*-Y* association will not perfectly represent the true A-Y effect unless it is zero (Fig. 2a).

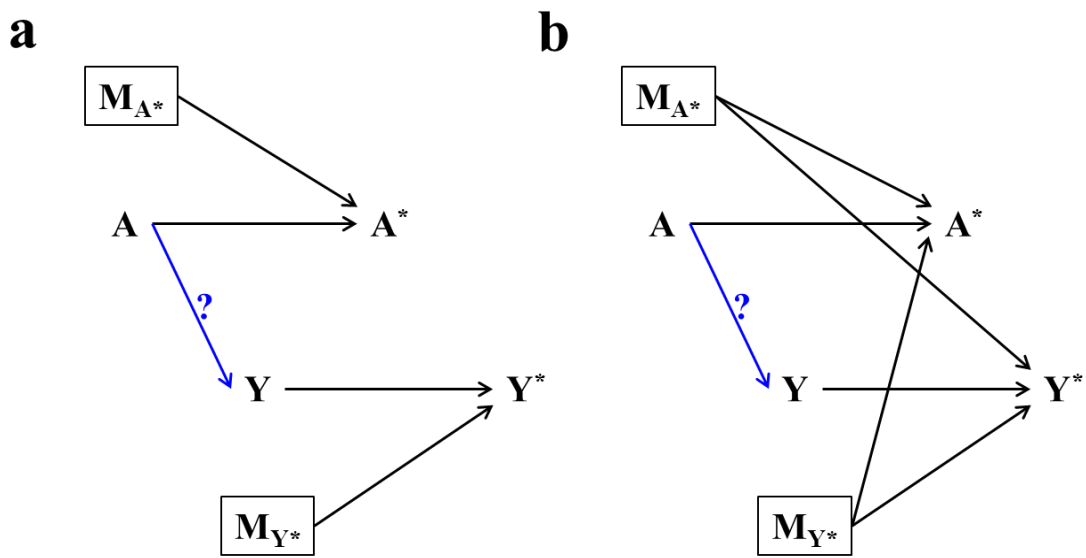

**Fig. 2 A directed acyclic graph for an etiological study between an exposure (A) and an outcome (Y) (a- an independent mechanism, b- a dependent mechanism)**

A/Y: true value of an exposure/an outcome; A*/Y*: measured version of A/Y; $M_{A*}$/$M_{Y*}$: candidate measurement systems for the value of A/Y; The arrow with a question mark near it means that the relationship between two variables at the ends of the arrow is under study.

In Fig. 2a, we can infer that the A*-Y* effect is actually zero since they are connected by no arrow. However, A* and Y* will be associated, as indicated by an open path from A* to Y* (A*←A→Y→Y*), where the exposure (A) itself acts as a confounder in essence if A does have an effect on Y. We call this as a substituted estimate [31,37], which represents our study interest. If A does not have an effect on Y, the substituted estimate would be zero.

Since the MB for the two singleton variables are represented by either of the paths, $M_{A*}$→A* and $M_{Y*}$→Y* respectively, as seen from Fig. 2a, Y* could not affect the MB

7 / 22

for A*, and A* could not affect the MB for Y*. Thus, the values of both A* and Y* are misclassified non-differentially if both are categorical variables (Fig. 2a).

Contrary to the measurement for a singleton variable, $M_{A*}$ may be associated with $M_{Y*}$ by any open path between them through any mechanism (for example, a common cause or any causality), thus the measurements for A and Y may be dependent (Fig. 2b, A*←$M_{A*}$---$M_{Y*}$→Y*). However, generally only one AVMS has been respectively applied for the measurements of the value of A and Y, these paths from A* through $M_{A*}$ and / or $M_{Y*}$ to Y* will all blocked anyway. For example, in the study of mother-to-infant transmission of chronic hepatitis B virus infection, the status of this infection in the mother-infant pairs may be tested with the same one kit, which blocks the path of A*←$M_{A*}$/$M_{Y*}$→Y* (at this time, $M_{A*}$=$M_{Y*}$). Consequentially, the possible association between $M_{A*}$ and $M_{Y*}$ will not change the characteristic of non-differential misclassification mentioned above.

It can be seen from the analysis above that the MBs for our substituted estimate are originated from both the singleton variables (A* and Y*) due to their corresponding measurement systems. Nevertheless, in order to satisfy the feasibility of etiological studies, we usually need to assume that the substituted estimate is a reasonable and feasible though biased alternative for the A-Y effect, and then used for causal inference. This is exactly the implications underlying all etiological studies.

## 2.2. The MB for the substituted estimate: the sources other than the measurement systems

Under the interdependency and dependency mechanism, only the measurement systems may non-differentially affect the MB for both the measured exposure (A*) and outcome (Y*). However, it is possible that extra factors will exert some influence on both measurements. Since both A and Y are exogenous variables when measured alone, we can reason that the additional MBs for any singleton variable except the effect of the measurement systems will influence either the effect of A* from A or that of Y* from Y. Since the A*-Y* association will be attributable to any of the following four causal mechanisms [38], we will elaborate on them separately under the independent mechanism of measurement for simplicity and a non-null effect firstly. Since generally only one AVMS was used for either of the exposure and the outcome, we will omit the AVMS from the following diagrams.

(1) Common causes between A* and Y*

In this mechanism, two new and closed paths are present, i.e., A→Y→Y*←⋯←C→⋯→A* and Y←A→A*←⋯←C→⋯→Y* (the symbol "⋯" represents all the single arrows on the path pointing to the same direction), which will not add new influence on the value of either A* or Y*. Thus, the values of both A* and Y* are mutually misclassified non-differentially if both are categorical variables (Fig. 3a).

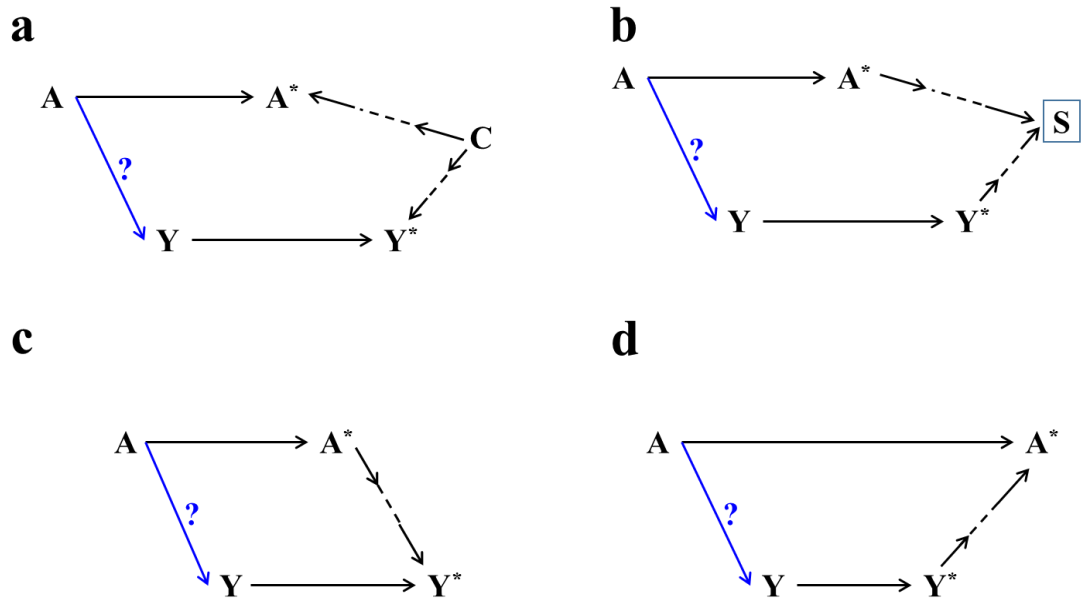

**Fig. 3 A directed acyclic graph for an etiological study between an exposure (A) and an outcome (Y) under four different mechanisms**

A/Y: true value of an exposure/an outcome; A*/Y*: measured version of A/Y; C: a confounder; S: a collider. The box around S means conditioning. The symbol "⋯" represents all the single arrows on the path pointing to the same direction. The arrow with a question mark near it means that the relationship between the two variables at the ends of the arrow is under study.

In the example of estrogen use and occurrence of endometrial cancer [39], since both estrogen use and endometrial cancer may increase the chance of vaginal bleeding (B), which can be identified by the woman herself and/or diagnosed by her doctor (B*). Diagnosed vaginal bleeding (B*) can increase the chance of the patient's clinical visits and examinations, which will in turn increase her chance of being diagnosed with endometrial cancer (Y*). This often-called detection bias can be explained by the extra confounding path irrelevant to our interest, A*←A→B→B*→Y* (Fig. 4). This mechanism is similar to that of dependency mechanism for the measurement systems or common causes between A* and Y* mentioned above. It should be emphasized here that it is the measured and known-to-the-woman intermediator outcome (B*), not its unmeasured and unknown one (B), that increases her chance of being diagnosed with endometrial cancer (Y*) directly.

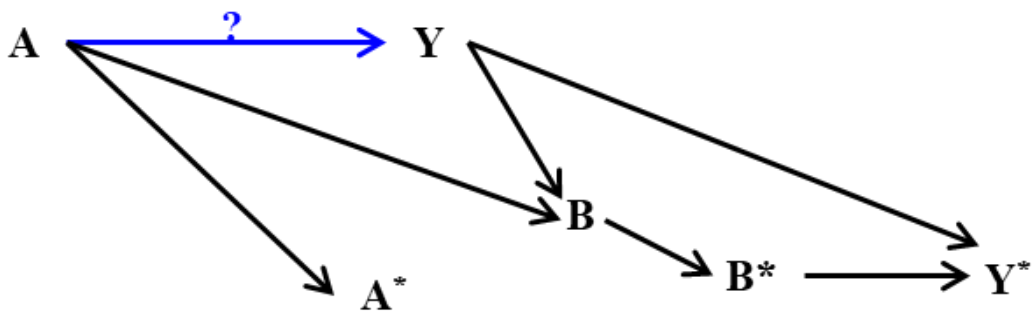

**Fig. 4 A directed acyclic graph for an etiological study between estrogen use (A) and endometrial cancer (Y)**

A: true status of estrogen use; A*: measured version of A; Y: true status of endometrial cancer; Y*: measured version of Y; B: true status of vaginal bleeding; B*: measured version of B. The arrow with a question mark near it means that the relationship between the two variables at the ends of the arrow is under study.

### (2) Common outcomes between A* and Y*

In this mechanism, two new paths are present again, i.e., A→Y→ Y*→⋯→S←⋯←A* and Y←A→A*→⋯→S←⋯←Y*. Both paths are closed by the collider S but will turn on by conditioning on S, thus the values of both A* and Y* will be affected, which will in turn result in mutually differential misclassifications if both are categorical variables

(Fig. 3b).

For example, when both the exposure (A) and the outcome (Y) are simultaneously measured, as in record-based studies, or historical cohort studies (Fig. 3b), A* and Y* are causally irrelevant. From this point of view, these studies are less likely to produce additional MB.

Since these studies often restrict to those subjects on the record (here, if S=1, the subjects were recorded previously; if S=0, the subjects were not), we are conditioning on S=1. This is the underlying mechanism of MB induced by selection bias, which can be explained by the extra open path irrelevant to our interest, A*→⋯→S←⋯←Y* (Fig. 3b). It occurs when those recorded are not comparable to those not recorded, which can further lead to a biased estimate of substituted estimate, or informed presence bias [40,41].

### (3) A causal relationship between A* and Y*

The situation can be somewhat different if A* and Y* has a causal relationship.

Firstly, if A* causally affects Y*, we will have two new paths, i.e., A→Y→Y*←⋯←A* and Y←A→A*→⋯→Y*. The former is closed, however the latter is open, thus the value of Y* is misclassified differentially depending on the value of A*, whereas A* is misclassified non-differentially depending on the values of Y* if both are categorical variables (Fig. 3c).

In the example of Hawthorne effect on the study of viral disease (A) and antibiotic prescription (Y) [42], children with viral disease diagnosed by their doctors (A*) participated in the study. This resulted in the increase of awareness among doctors since they are surveyed and audiotaped, which in turn lowered their report of antibiotic prescription (Y*) to the investigators. This is often called as a causality mechanism. Again it should be emphasized here that it is the children diagnosed by the doctors (A*) which has changed the doctors' report of their prescribing behavior (Y*).

Secondly, when the causal relationship between A* and Y* is reversed, which we may frequently see in case-control studies, we will also have two new paths, i.e., A→Y→Y*→⋯→A* and Y←A→A*←⋯←Y*. The former is open, but the latter is closed, thus the value of Y* is misclassified non-differentially depending on the value of A*, but the value of A* is misclassified differentially depending on the values of Y* if both are categorical variables (Fig. 3d).

This mechanism exemplifies the recall bias we frequently encounter in case-control studies. In a study of mothers who delivered a baby with congenital malformation, they may overestimate or underestimate their previous exposure (A*) after their infants are diagnosed (Y*). This is often called a reverse causality mechanism, which can be explained by the extra open path irrelevant to our interest, Y*→⋯→A* (Fig. 3d). Again, it should be emphasized here that it is the measured and known-to-the-woman outcome (Y*) that affects the measured exposure (A*), not its unmeasured and unknown outcome (Y) that affects the exposure (A). Thus, reverse causality is a fuzzy and unnecessary concept that is actually defined at the measured level, but it is often misrepresented as Y→A in the DAGs reported in many literatures. However, cause and effect will never be reversed!

When we further refer to a dependent mechanism none of the above four mechanisms will change our conclusions on the classification, direction, or their susceptibility to interdependency for both A* and Y* (Tab. 1). For a null effect, all the mechanisms would result in a non-differential misclassification.

**Tab. 1 The differential and non-differential misclassification scheme for a measured exposure (A\*) and a measured outcome (Y\*) under a null and non-null A-Y effect**

| Mechanisms under a null or non- null effect | The value of Y* misclassified depending on the values of A* | The value of A* misclassified depending on the values of Y* |
| --- | --- | --- |
| Null effect | | |
| Common causes between A* and Y* | Non-differentially | Non-differentially |
| Common outcomes between A* and Y* | Non-differentially | Non-differentially |
| A* causally affects Y*, directly or indirectly | Non-differentially | Non-differentially |
| Y* causally affects A*, directly or indirectly | Non-differentially | Non-differentially |
| Non-null effect | | |
| Common causes between A* and Y* | Non-differentially | Non-differentially |
| Common outcomes between A* and Y* | Differentially | Differentially |
| A* causally affects Y*, directly or indirectly | Differentially | Non-differentially |
| Y* causally affects A*, directly or indirectly | Non-differentially | Differentially |

**Notes**

A: an exposure; A*: the measured version of A; Y: an outcome; Y*: the measured version of Y.

The results are the same for the mechanisms of both independence and dependence

It can be seen from the above that the presence of MB for the substituted estimate resulted from a redundant A*-Y* association that is irrelevant to our research interest. Irrespective of the type (either quantitative or categorical) or dimension (two or multiple) of the variables, this association must result from any combination of the following three intrinsically mutually exclusive pairs of causal structures, i.e., independence and dependence, common causes and common outcomes, and causality and "reverse causality".

For misclassification, these structures will be helpful in the judgement of the classification and direction, two characteristics that do not interfere with each other for misclassification bias. The main points are as follows: (1) For a null effect or a common-cause structure, both the values of A* and Y* show a non-differential misclassification. (2) For a common-outcome structure, both the values of A* and Y* show a bidirectionally differential misclassification. (3) If A*-Y* has a causal relationship, either A*→⋯→Y* or Y*→⋯→A*, the misclassification will be unidirectional, i.e., A* (or Y*) will only result in a unidirectional differential misclassification of Y* (or A*); but Y* (or A*) will only result in a non-differential misclassification of A* (or Y*).

### Discussion

In this paper, we used DAGs to propose a new structure for measuring one singleton variable, and then extend it into clarifying causal relationships between two variables. The MB of a singleton variable often roots in a selection of a practically imperfect measurement system. This imperfection, in combination with any redundant association between a measured exposure and a measured outcome, will result in the MB of an effect estimate.

Theoretically, MB is the difference between the true value and its measured version for both a singleton variable and the relationship between variables. Based on the I/O device-like measurement process for a singleton variable, we propose a new MB-DAG that combines prior knowledge, causal reality and selection process of the measurement systems. Our new structure has the following advantages. Firstly, the explicit definition of the measure system as a device-like way has highlighted its input-output role, and enriched the meaning of $U_{A*}$. Secondly, it will aid in our understanding of different types of MB. A measurement system with some specific imperfections will lead to its corresponding bias. For example, the researchers' careless observing will result in observer bias; the subjects' imperfect reporting may result in reporting bias; a non-standardized method will also result in some bias if there are some violations in the standardized procedure. Thirdly, we can quickly apply this idea into the validity studies among different measurement systems.

When this new DAG is extended to causal relationships between an exposure (A) and an outcome (Y), the MB for each of their measured version, A* and Y*, will result in an inevitable MB for the substituted estimate of the A-Y effect. This has shifted our focus to the assurance of the correctness of this substituted estimate.

Generally, there are two different sources for this MB, one from the measurement system and the other from any redundant association between A* and Y*. Irrespective of the type (quantitative or categorical) of these two variables, the MB of the substituted estimate roots in the following three mutually exclusive pairs of causal structures, (1) independence and dependence, (2) common causes and common outcomes, (3) causality and "reverse causality." These structures have been illustrated in details and with illustrations. The final MB in any scientific study will be any combination of these structures above.

These mechanisms further dissect graphically the scheme of misclassification and direction when the variables are categorical. Given a specific measurement for both A* and Y*, a non-differential misclassification will arise for the scenario of a null A-Y effect, or that of a common cause for both A* and Y*; and a bidirectionally differential misclassification will be introduced for the scenario of a common outcome for both A* and Y*. However, if A* and Y* has a causal relationship, the misclassification will be unidirectional. A*/Y* will only result in a differential misclassification of Y*/A*, but Y*/A* will only result in a non-differential misclassification of A*/Y*. Under this idea, the "bias toward the null" maxim for categorical variables will be quickly refuted [9]. To our knowledge, this is the first time to clarify the direction of misclassification from causal graph.

We illustrated with several examples to clarify that measurement will lead to the continuum of causality from real world to human thinking. As in our examples, it is the measured version of the true value resulted in the consequence that the stakeholders could make any psychological or behavioral change themselves [9], which in turn lead to their final measured companions, i.e., the measured exposure or outcome. For example, women with diagnosed vaginal bleeding increased clinical visits which had resulted in more ascertainment of endometrial cancer; doctors' participation in the study changed their behaviors in prescribing antibiotics; women who delivered a baby with congenital malformation would overestimate or underestimate their previous exposure.

MB has been largely ignored [43], but it does matter. Thus, it is essential to obtain a correct measurement for any singleton variable as possible as we can. Otherwise, we have to turn to effect restoration when validation data are available [35,44,45]. Our new DAG is helpful in understanding and directing practical studies when referring to MB.

In conclusion, measurement is a complex causal process, which touches on the continuum of causality from real world to human thinking. As MB will be present inevitably, we can only assure a correct estimate of the measured version; however, it may still be far away from the truth if effect restoration has not been undertaken. DAG is a powerful tool in clarifying the structures and mechanisms of MB.

**Acknowledgments**

We would like to thank Dr. Qiang Xia, Bureau of HIV/AIDS Prevention and Control, New York City Department of Health and Mental Hygiene, for his valuable comments on this paper. This work was supported by National Natural Science Foundation of

China [81373065, 81773490] and The National Key Research and Development Program of China [2017YFC1200203].19 / 22

Science+Business Media; 2009.